\documentstyle[12pt]{article}

\begin{document}

\begin{center}
{\bf HELICAL VERSUS FUNDAMENTAL SOLITONS IN OPTICAL FIBERS}
\end{center}

\bigskip

\begin{center}
Boris A. Malomed\footnote{%
e-mail: malomed@eng.tau.ac.il}

Department of Interdisciplinary Studies, Faculty of Engineering, Tel Aviv
University, Tel Aviv 69978, Israel
\end{center}

\bigskip

\begin{center}
G.D. Peng and P.L. Chu

Optical Communications Group, School of Electrical Engineering and
Telecommunications, University of New South Wales, Sydney 2052, Australia
\end{center}

\newpage

\begin{center}
{\large {\bf ABSTRACT}}
\end{center}

We consider solitons in a nonlinear optical fiber with a single polarization
in a region of parameters where it carries exactly two distinct modes, viz.,
the fundamental one and the first-order helical mode. From the viewpoint of
applications to dense-WDM communication systems, this opens way to double
the number of channels carried by the fiber. Aside from that, experimental
observation of helical (spinning) solitons (that can be launched and
detected, using helicity-generating phase masks) and collisions between them
and with fundamental solitons in (ordinary or {\em hollow}) optical fibers
is an issue of fundamental interest, especially because it has been very
recently found that spatiotemporal spinning solitons in bulk optical media
with various nonlinearities are unstable. We introduce a system of coupled
nonlinear Schr\"{o}dinger equations for fundamental and helical modes,
computing nonstandard values of the cross-phase-modulation coupling
constants in it, and investigate, analytically and numerically, results of
``complete'' and ``incomplete'' collisions between solitons carried by the
two modes. We conclude that the collision-induced crosstalk is partly
attenuated in comparison with the usual WDM system, which sometimesamay
be crucially important, preventing merger of the colliding
solitons into a breather. The interaction between the two modes is found to
be additionally strongly suppressed in comparison with that in the WDM
system in the case when a dispersion-shifted or dispersion-compensated fiber
is used.

\newpage

\section{Introduction}

A commonly adopted approach to the description of nonlinear light
propagation in optical fibers is based on separation of the transverse modal
structure, that may be described in the linear approximation, and slow
longitudinal and temporal evolution of the signal's envelope, which is
essentially affected by the temporal dispersion and Kerr nonlinearity. The
analysis following this way ends up with the derivation of the nonlinear Schr%
\"{o}dinger (NLS) equation for the envelope \cite{Agr,HK}.

Usually, both experimental and theoretical studies of the soliton
propagation are confined to the case when parameters of the fiber admit a
single (fundamental) transverse mode, because in a multimode fiber an
initial pulse excites different modes in an uncontrollable fashion. However,
using well-known data for the fibers of the simplest step-index type \cite
{Love}, it is easy to find that a situation with exactly two modes
takes place when the standard {\it waveguide parameter} 
\begin{equation}
V\equiv k\rho \sqrt{n_{{\rm co}}^{2}-n_{{\rm cl}}^{2}},  \label{V}
\end{equation}
where $k$, $\rho $ and $n_{{\rm co,cl}}$ are, respectively, the carrier
wave's propagation constant, core radius, and the refractive index in the
core and cladding, takes values 
\begin{equation}
2.405<V<3.832\,\,.  \label{existence}
\end{equation}
For instance, in the case of the standard carrier wavenumber $\lambda
=1.54\,\mu $m admitting the soliton propagation in optical fibers, and
taking the usual value $n_{{\rm co}}-n_{{\rm cl}}=0.01$, the interval (\ref
{existence}) implies $3\,\mu $m $<\rho <4.75\,\mu $m, i.e., quite realistic
values of the core's radius. 

Inside the interval (\ref{existence}), the fiber carries a fundamental mode
(FM) and the first helical mode (HM). The transverse structure of the
latter one is described by the expressions $J_{1}(Ur)\exp (\pm i\theta )$ in
the core and $K_{1}(Wr)\exp (\pm i\theta )$ in the cladding, where $J_{1}$
and $K_{1}$ are the standard cylindrical functions, $U$ and $W$ are the
associate waveguide parameters defined in the usual way \cite{Love}, and ($r$%
,$\theta $) are the polar coordinates in the fiber's cross section. In this
case, one is actually dealing with a set of three coexisting modes, as there
are two degenerate HMs with the helicities $S=\pm 1$. Note also that,
because the physical fields are proportional to the real part of the complex
expressions, the presence of the multiplier $\exp (\pm i\theta )$ means that
HM solitons are {\em spinning} in the course of the propagation along the
fiber.

As neither higher-order radial (nonhelical) mode, nor any HM with a
helicity $\,>1$ exists in the interval (\ref{existence}), the two-mode
situation is controllable: a light pulse with zero helicity can
excite solely FM, while excitation of HM is possible by a pulse that carries
the necessary helicity. A light beam can be lent helicity, passing it
through a specially designed phase mask, which is quite feasible in a real
experiment, see, e.g., Ref. \cite{Petrov}. Due to their distinct topological
nature, FM and HM do not linearly mix, provided that the fiber remains
circular. As is well known, it is easy to fabricate a long silica fiber
whose deviation from circularity is negligible. Fiber's bending will induce
no linear mixing either, providing that the bending radius is much larger
than the wavelength. In this work, however, we do not discuss exact
limitations on the deviation of the fiber from the circularity and similar
details. Instead, we focus on principal issues, such as collisions between
solitons carried by FM and HM.

It is necessary to distinguish between the fundamental and helical solitons
at the receiver end of the fiber. In an experiment with a single or a few
copropagating carrier frequencies, this is quite simple, as the fundamental
and helical modes have an appreciable difference in their propagation
constants (see below), thus the two types of the solitons can be
discriminated between by means of a simple wavelength filter. Besides that,
there is
a possibility to create a ``helicity filter'', which would also
work in the case of a multi-channel WDM system. Indeed, if it is known that
two species of the solitons in the fiber have the helicities $S=0$ and $S=+1$,
at the receiver end the incoming signal can be passed through a phase mask
that adds extra helicity
$\Delta S=-2$. Then, the arriving FM and HM solitons will change
their helicities to $-2$ and $-1$, respectively, so that only the latter one
will survive, as in the selected parametric region the modes with $S=\pm 2$
do not propagate in the fiber. On the other hand, passing the incoming
signal through the phase mask adding $\Delta S=+1$ will transform the former 
$S=0$ soliton into a propagating $S=+1$ pulse, while the former $S=+1$
soliton will have $S=+2$, hence it will not be able to propagate.

Thus, launching solitons independently in each of the two modes, one can
implement a two-channel system inside the core. Note that standard elements
of the fiber communication systems, such as amplifiers and guiding filters 
\cite{HK}, will act in essentially the same way on the solitons in both
modes (although the gain coefficients of an Er-doped or Raman fiber
amplifier may differ for the two modes, depending, e.g., on the density
distribution of the doping atoms in the fiber's transverse plane). Moreover,
if one starts with a WDM (wavelength-division-multiplexed) multichannel
system already implemented in the fiber, one can {\em double} the number of
the channels by means of this two-mode scheme (we demonstrate below that it
is not really possible to triple the number of the channels, using two HMs
with the opposite helicities). The feasibility of such a ``mode-division''
channel doubling may be quite important, as it has been demonstrated that
doubling by means of the polarization-division multiplexing is incompatible
with WDM \cite{Elgin}. Indeed, while the polarization of the soliton can be
easily changed by various imperfections of the system, the mode's helicity
is expected to be robust, as it is a {\it topological invariant}. Note that
we do not consider 
the polarization structure of the modes, assuming that
either they belong to one polarization, or (more realistically for the
applications) the polarization can be effectively averaged out. Thus, our
helical mode has nothing in common with the circular polarization.

Due to the Kerr nonlinearity, the linearly orthogonal solitons borne by the
two modes interact via the cross-phase modulation (XPM). The main technical
objective of this work is to study XPM-induced effects of collisions between
the solitons. As is well known, the collisional crosstalk is the most
fundamental problem in the soliton-based multichannel communication systems,
see. e.g., Refs. \cite{coll}~-~\cite{KMJ}.

It should be stressed that, while the application of the proposed
mode-division doubling to WDM soliton communication systems is not
straightforward, as some technical problems remain to be resolved,
experimental observation of narrow {\em subpicosecond} helical (spinning)
solitons and their collisions between themselves and with fundamental
solitons in relatively short optical fibers is a problem of fundamental
physical interest by itself. A combination of the above-mentioned
helicity-generating phase masks with the well-developed experimental
technique admitting, e.g., direct observation of the polarization structure
of subpicosecond solitons in short fibers \cite{Silberberg} should make the
observation of the helical solitons and their collisions quite feasible. An
additional interest to the latter problem is lent by the fact that a similar
object in a bulk optical medium, viz., a {\it spinning light bullet}
(spatiotemporal soliton), has been very recently found to be unstable in
models with various nonlinearities \cite{spinning}. Therefore, the optical
fiber may be a unique medium in which the existence of a stable {\em spinning%
} temporal soliton may be possible.

It is also noteworthy that the helical soliton, whose local intensity
vanishes at the central point of the fiber's cross section, may be a natural
object to exist in {\em hollow }nonlinear optical fibers, which have
recently attracted a lot of attention (and where, incidentally, very narrow
solitary pulses are quite possible), see, e.g., the works \cite{hollow} and
references therein. An interesting issue is a possibility to select
parameters of the hollow fiber so that it would support solely a helical
mode, which would then play the role of the fundamental one.

The paper is organized as follows. In section 2, a system of coupled NLS
equations for the three modes, FM with $S=0$ and two HMs with $S=\pm 1$, is
obtained. It has nonstandard values of the XPM coupling constants. In
section 3, collisions between solitons carried by the fundamental and
helical modes is studied analytically, by means of the perturbation theory.
The perturbative treatment applies to the case when the colliding solitons
pass through each other quickly enough. In section 4, an example of direct
numerical simulations of the collision between the FM\ and HM solitons,
illustrating a difference from the collision between FM solitons in the
usual WDM system, is displayed. The difference may be crucial: the colliding
solitons merge into a breather in the usual system, but survive the
collision when they belong to the different modes. The paper is concluded by
section 5.

\section{The Model}

A normalized system of coupled nonlinear Schr\"{o}dinger (NLS) equations for
the interacting modes can be derived by means of a standard asymptotic
procedure \cite{Agr}, 
\begin{eqnarray}
i(u_{0})_{z}+ic_{0}\left( u_{0}\right) _{\tau }+k_{0}u_{0}-\frac{1}{2}\beta
_{0}(u_{0})_{\tau \tau }+\left( |u_{0}|^{2}+2\gamma _{0}|u_{+}|^{2}+2\gamma
_{0}|u_{-}|^{2}\right) u_{0}\, &=&\,0\,,  \label{0} \\
i(u_{+})_{z}+ic_{1}\left( u_{+}\right) _{\tau }+k_{1}u_{+}-\frac{1}{2}\beta
_{1}(u_{+})_{\tau \tau }+\left( |u_{+}|^{2}+2|u_{-}|^{2}+2\gamma
_{1}|u_{0}|^{2}\right) u_{+}\, &=&\,0\,,  \label{+1} \\
i(u_{-})_{z}+ic_{1}\left( u_{-}\right) _{\tau }+k_{1}u_{-}-\frac{1}{2}\beta
_{1}(u_{-})_{\tau \tau }+\left( |u_{-}|^{2}+2|u_{+}|^{2}+2\gamma
_{1}|u_{0}|^{2}\right) u_{-}\, &=&\,0\,  \label{-1}
\end{eqnarray}
(in the case of extremely narrow solitons, 
well-known higher-order terms 
\cite{Agr,HK} should be added to the system). We here consider the most general case, when two HMs with the helicities $\pm 1$, represented by the envelopes 
$u_{\pm }$, interact with the zero-helicity
FM $u_{0}$; $\beta _{0}$ and $\beta _{1}$
are the corresponding mode-dependent dispersion coefficients (see below), $%
c_{0,1}$ and $k_{0,1}$ are the group-velocity and propagation-constant
shifts of the two modes (these characteristics are also mode-dependent \cite
{Love}), and the effective XPM coefficient $\gamma _{0}$ and $\gamma _{1}$
are given by the properly normalized overlapping integrals between FM and
HM. Using known expressions for the transverse modal functions of the
step-index fiber \cite{Love}, we have calculated them numerically. In Fig.
1a, we display $\gamma _{0}$ and $\gamma _{1}$ vs. the waveguide parameter (%
\ref{V}).

Note that Eqs. (\ref{0})~-~(\ref{-1}) do not contain four-wave mixing (FWM)
terms. Some of them might originate from the terms $\sim u_{0}^{2}\left(
u_{\pm }^{\ast }\right) ^{2}$ and its complex conjugate in the model's
Hamiltonian density. However, the full expressions to be inserted into the
Hamiltonian are multiplied by the modal angular dependences $\exp (\mp
2i\theta )$, hence they will give zero upon the angular integration. Another
possible source of FWM terms in Eqs. (\ref{0})~-~(\ref{-1}) could be the
term $\sim u_{0}^{2}u_{+}^{\ast }u_{-}^{\ast }$ and its complex conjugate in
the Hamiltonian density. In these expressions, the angular dependence
cancels out, hence the angular integration will not nullify them. However,
the corresponding terms in Eqs. (\ref{0})~-~(\ref{-1}) will be rapidly
oscillating in $z$ because of the difference between the propagation
constants $k_{0}$ and $k_{1}$. A straightforward consideration yields an
estimate for the relative wavenumber mismatch between FM and HM in the
region of interest, $|k_{1}-k_{0}|/k\approx 0.4\left( n_{{\rm co}}-n_{{\rm cl%
}}\right) $. Taking the same estimate for the refractive index
difference, $n_{{\rm co}}-n_{{\rm cl}}\approx 0.01$, as above, we conclude
that $|k_{1}-k_{0}|/k\sim 0.005$, which corresponds to the beat length $\sim
200$ wavelengths. As it is disparately small in comparison with any
propagation distance relevant to the solitons, all the FWM terms may be
neglected.

As for the dispersion coefficients in Eqs. (1)~-~(3), their parts accounted
for by the waveguide geometry can also be calculated for the two modes on
the basis of the data available from the linear-propagation theory \cite
{Love}. The result of the calculation is shown in Fig. 1b. It is noteworthy
that the waveguide-geometry part of $\beta _{0}$ changes its sign. One
should, however, keep in mind that the full dispersion also contains a
material (bulk) contribution, that may be essentially larger than that
displayed in Fig. 1b. 

Thus, the analysis of the interaction between
solitons must admit {\em different} (but both negative, i.e., anomalous
\cite{Agr}) effective dispersions 
$\beta _{0}$ and $\beta _{1}$ in Eqs. (\ref{0})~-~(\ref{-1}). 
Together with the nonstandard values of the XPM coefficients 
$\gamma _{0}$ and $\gamma _{1}$, these features constitute an essential
mathematical difference of the present model from a usual three-channel WDM
one (see, e.g., Ref. \cite{Abl}).

The fundamental and helical modes are also characterized by a difference in
their group velocities, which plays a crucially important role in the
analysis of soliton-soliton collisions. Continuing the above estimates of
the physical parameters for the present case, we obtain 
\begin{equation}
\left( \delta v_{{\rm gr}}\right) _{{\rm mode}}/v_{{\rm gr}}\sim 5\cdot
10^{-6}  \label{mode}
\end{equation}
for the relative group-velocity difference between the modes. As concerns
the possibility to use the mode-division doubling of the channels in the WDM
system, it is relevant to mention that, in the WDM system implemented in the
standard telecommunications fiber with the dispersion $\beta \sim -20$ ps$%
^{2}$/km at $1.54$ $\mu $m, the relative group-velocity mismatch between the
adjacent channels is estimated to be 
\begin{equation}
\left( \delta v_{{\rm gr}}\right) _{{\rm WDM}}/v_{{\rm gr}}\sim 10^{-2}\cdot
\left( \delta \lambda /\lambda \right) \,,  \label{WDM}
\end{equation}
$\delta \lambda $ being the wavelength separation between the channels. The
case of practical interest is $\Delta \lambda \sim 1$ nm, hence we conclude
that the relative group-velocity differences (\ref{mode}) and (\ref{WDM})
are on the same order of magnitude. On the other hand, in the
dispersion-shifted (DS) or dispersion-compensated (DC) fibers, the effective
value of the dispersion is much smaller than the above-mentioned value $-20$
ps$^{2}$/km, hence in these cases the inference is that the corresponding
WDM relative difference is negligible as compared to that between the
fundamental and helical modes, 
\begin{equation}
\left( \delta v_{{\rm gr}}\right) _{{\rm WDM}}^{{\rm (DS/DC)}}/v_{{\rm gr}%
}\ll \left( \delta v_{{\rm gr}}\right) _{{\rm mode}}/v_{{\rm gr}}.
\label{compare}
\end{equation}

For typical solitons to be used in telecommunications, with the temporal
width $T\sim 10$ ps, the above estimate $\left( \delta v_{{\rm gr}}\right) _{%
{\rm mode}}/v_{{\rm gr}}\sim 5\cdot 10^{-6}$ implies that a collision
between the FM and HM solitons takes place at a propagation distance $z_{%
{\rm coll}}\sim T\left( \delta v_{{\rm gr}}\right) _{{\rm mode}}/v_{{\rm gr}%
}^{2}\sim 500$ m, which is much shorter than any soliton's length scale.
This circumstance allows us to treat the XPM-mediated interaction as a small
perturbation in the course of the fast passage of one soliton through the
other, as it was done in other contexts in Refs. \cite{Abl,KMJ}. Note that
in the laboratory experiments with subpicosecond solitons, the collision
length may be $\,\,_{\sim }^{<}\,50$ m, implying that the experimental study
of the collisions should be quite possible.

However, there is no group-velocity difference between the two HMs $u_{\pm }$%
, hence the collision distance for the corresponding solitons may be very
large, giving rise to a strong crosstalk between them. Moreover, the
collision between two solitons with the helicities $S=\pm 1$ may result in
their annihilation or transformation into a pair of $S=0$ solitons, while,
due to the conservation of the topological invariant, the collision between
the solitons with $S=0$ and $S=1$ is expected to be much closer to an
elastic one. In view of this, it makes sense to assume only the doubling of
the number of channels by means of the ``mode-division multiplexing'' (i.e.,
to use only one HM) in the context of the WDM systems, but not tripling,
that might seem possible due to the existence of two HMs with $S=\pm 1$.
Irrespective of that, a study of collisions between the solitons with $S=+1$
and $S=-1$ is a challenge for experiments with narrow solitons in
optical fibers.

\section{Analytical Treatment of Soliton Collisions}

Proceeding to the perturbative analysis of the collision between the
solitons carried by FM and HM, we should take into regard that, in view of
the asymmetry between Eqs. (\ref{0}) and (\ref{+1}), (\ref{-1}), the FM and
HM solitons may have {\em different} widths, $T_{0}$ and $T_{1}$. This
circumstance makes it technically impossible to base the perturbative
treatment of the collision on the exact unperturbed soliton waveforms of the 
${\rm sech}$ type, as the corresponding overlapping integrals will be
intractable. The only possibility to develop an efficient perturbation
theory is to use, as the zero-order approximation, the Gaussian {\it ansatz}
for the unperturbed soliton solutions to the uncoupled equations (\ref{0})
and (\ref{+1}), 
\begin{equation}
u_{l}^{(0)}(z,\tau )=A_{l}\exp \left( iK_{l}\,z-\frac{(\tau -t_{l})^{2}}{%
2T_{l}^{2}}\right) ,\;\frac{dt_{l}}{dz}=c_{l};\;l=0,1\,,  \label{Gauss}
\end{equation}
where a relation between the amplitude and width of the soliton can be found
by means of the variational approximation \cite{Dan}, 
\begin{equation}
A_{l}^{2}=\sqrt{2}|\beta _{l}|/T_{l}^{2}  \label{Amplitudes}
\end{equation}
(the propagation constants $K_{l}$ will not be needed here). In fact, a
difference between the approximate soliton's shape given by Eq. (\ref{Gauss}%
) and the exact ${\rm sech}$ shape is fairly small, see, e.g., Fig. 5 in
Ref. \cite{Dan}.

A soliton moving in the given reference
frame is obtained from Eq. (\ref{Gauss}) as its Galilean transform, 
\begin{equation}
u_{l}(z,\tau )=u_{l}^{(0)}(z,\tau -t_{l}(z))\,\exp \left( -i\omega _{l}\tau
+iqz\right) \,,  \label{Galileo}
\end{equation}
where $\omega _{l}$ is an arbitrary transform-generating frequency shift,
the propagation-constant shift $q$ is not essential, and (cf. Eq. (\ref
{Gauss})) 
\begin{equation}
\frac{dt_{l}}{dz}=c_{l}-|\beta _{l}|\omega _{l}\,.  \label{omega}
\end{equation}

If the XPM term in Eqs. (1)~-~(3) is, effectively, a small perturbation (in
the case of a fast collision, see above), the collision between the solitons
may be described as that between two quasiparticles interacting through an
effective potential. Following the lines of the analysis developed for
similar problems earlier \cite{Abl,KMJ}, it is straightforward to derive the
following perturbation-induced evolution equations for the solitons'
frequency shifts: 
\begin{equation}
\frac{d\omega _{l}}{dz}+\frac{4|\beta _{l}|\gamma _{l}}{T_{1-l}\sqrt{%
T_{0}^{2}+T_{1}^{2}}}\cdot \frac{d}{dt_{l}}\,\exp \left[ -\frac{%
(t_{1}-t_{0})^{2}}{2(T_{0}^{2}+T_{1}^{2})}\right] =0\,,  \label{dynamics}
\end{equation}
where Eq. (\ref{Amplitudes}) was used to eliminate the amplitudes in favor
of the widths $T_{l}$ (recall that $\gamma _{l}$ are the relative XPM\
coupling constants in Eqs. (\ref{0})~-~(\ref{-1})). Combining Eqs. (\ref
{dynamics}) with Eqs. (\ref{omega}) and assuming, in the first
approximation, $T_{l}={\rm \,const}$ furnish a closed dynamical system
governing the evolution of the temporal positions $t_{l}$ of the two
solitons.

To further apply the perturbation theory to Eq. (\ref{dynamics}), we recall
that, according to the estimates obtained above, the difference of the
inverse group velocities, $c\equiv c_{1}-c_{0}$, is, effectively, a large
parameter. Hence, in the lowest-order approximation, one may set $%
t_{1}-t_{0}\approx cz$ in the argument of the exponential in Eq. (\ref
{dynamics}), thus strongly simplifying the equation: 
\begin{equation}
\frac{d\omega _{l}}{dz}=\frac{4(-1)^{l}|\beta _{l}|\gamma _{l}}{cT_{1-l}%
\sqrt{T_{0}^{2}+T_{1}^{2}}}\cdot \frac{d}{dz}\,\exp \left[ -\frac{(cz)^{2}}{%
2(T_{0}^{2}+T_{1}^{2})}\right] \,.  \label{derivative}
\end{equation}

To proceed, it is necessary to specify a type of the collision to be
considered. One should distinguish between ``complete'' and ``incomplete''
collisions \cite{KMJ}. In the former case, the solitons are, originally, far
separated; in the course of the interaction, the faster soliton catches up
with the slower one and passes it. In the first approximation, the complete
interaction does not give rise to a net frequency shift (a change of the
frequency would be tantamount to a change of the soliton's velocity,
according to Eq. (\ref{omega})), as the integration of the right-hand side
of Eq. (\ref{derivative}) from $z=-\infty $ to $z=+\infty $ yields zero.
However, finding a nonzero {\em instantaneous} frequency shift from Eq. (\ref
{derivative}), inserting it into Eq. (\ref{omega}), and integrating the
latter equation yield a nonzero collision-induced {\em position} shift $%
\delta t_{l}$ of the soliton's center, which is the main effect of the
complete collision. A final result can be conveniently written as a relative
position shift, normalized to the soliton's temporal width: 
\begin{equation}
\frac{\delta t_{l}}{T_{l}}=(-1)^{l-1}\cdot 4\sqrt{2\pi }\,\frac{\beta
_{0}\beta _{1}}{c^{2}T_{0}T_{1}}\,\gamma _{l}.  \label{complete}
\end{equation}

An ``incomplete'' collision takes place if the solitons are essentially
overlapped at the initial point, $z=0$. This kind of the collision is more
significant, as it gives rise to a nonzero net frequency shift $\delta
\omega _{l}$ (hence, to a velocity shift too). The most important
(dangerous) case is
that when centers of the colliding solitons exactly coincide at $z=0$. In
this case, $\delta \omega _{l}$ is found by straightforward integration of
Eq. (\ref{derivative}) from $z=0$ to $z=+\infty $. The result can be
presented in a more natural form, multiplying the net frequency shift by the
soliton's temporal width (i.e., normalizing the frequency shift to the
soliton's spectral width): 
\begin{equation}
T_{l}\delta \omega _{l}=4(-1)^{l-1}\frac{|\beta _{l}|T_{l}}{cT_{1-l}\sqrt{%
T_{0}^{2}+T_{0}^{2}}}\,\gamma _{l}\,.  \label{incomplete}
\end{equation}

The only difference of Eqs. (\ref{complete}) and (\ref{incomplete}) from
similar results for the usual WDM system are the specific XPM coefficients $%
\gamma _{l}$, which are $\equiv 1$ in the usual case. The most promising
range for the applications is around $V=3.6$ (Fig. 1a), which gives 
\begin{equation}
\gamma _{0}\approx 0.98,\,\gamma _{1}\approx 0.62.  \label{gammas}
\end{equation}
This implies that the crosstalk between the FM and HM solitons is attenuated
by the factor $0.62$ for the HM soliton, as compared to the usual WDM
system, while for the FM soliton the crosstalk strength is not different
from that in the usual system (taking, instead, the values around $V=2.4$
(Fig. 1a), we will get small $\gamma _{1}\approx 0.25$, but large
$\gamma _{0}\approx 1.66$).

Thus, if the set of the HM and DM modes is used to double the number of the
channels in the WDM system, we conclude that the crosstalk for the HM
solitons due to their collisions with the FM ones in any channel is
attenuated, against the usual strength, by the above-mentioned factor 
$\approx 0.62$.

Note that the conclusion concerning the comparison with the WDM\ crosstalk
pertains to the case when the WDM\ system is realized in the standard
telecommunications fiber: as it was concluded above, in this case the
group-velocity difference between the fundamental and helical modes is on
the same order of magnitude
as the group-velocity mismatch between adjacent WDM channels, 
see Eqs. (\ref{mode}) and (\ref{WDM}). On the
contrary to this, in the DS or DC fiber link the group-velocity difference
between the FM and HM channels is much larger than that between the WDM
ones, hence the FM-HM crosstalk is much weaker than between the WDM
channels, according to Eqs. (\ref{complete}) and (\ref{incomplete}). 

\section{Numerical Simulations of the Collision}

Direct simulations of the soliton collisions within the framework of Eqs. (%
\ref{0}) and (\ref{+1}) for the two modes demonstrate that, although the
above-mentioned frequency-shift-attenuation factor $0.62$ is not really
small, sometimes it may be important. In Fig. 2a, we display an
example of a disastrous {\em incomplete} collision in the usual WDM system,
which leads to a merger of the solitons into a {\it breather} (the
simulations of the complete collision at the same values of the parameters
shows that it is fairly mild, producing only small position shifts of the
solitons). Replacing the usual XPM coefficients $\gamma \equiv 1$ by those
for the FM-HM collision, given by Eq. (\ref{gammas}), we find that the same
solitons {\em survive} the incomplete collision (Fig. 2b). 

\section{Conclusion}

We have considered solitons in a nonlinear optical fiber in a parametric
region where the fiber supports exactly two distinct modes, the fundamental
one and the first helical mode, which allows one to double the number of
soliton channels in the fiber. We have introduced a system of coupled NLS
equations for the two modes and computed nonstandard values of the relative
XPM coupling constants in it. Then, we investigated analytically and
numerically results of both ``complete'' and ``incomplete'' collisions
between solitons carried by the helical and fundamental modes, concluding
that the crosstalk effects are partly attenuated for the helical solitons.
The crosstalk between the two modes is found to be additionally 
suppressed in comparison with that in the WDM system in the case when the
latter system is realized in a dispersion-shifted or dispersion-compensated
fiber. Irrespective of the possible applications, experimental observation
of narrow helical (spinning) temporal solitons in ordinary or hollow optical
fibers is a challenge, especially because it has been recently demonstrated
that spinning spatiotemporal solitons (``light bullets'') are unstable in
bulk media.

\section*{Acknowledgement}

B.A.M. appreciates a support from the School of Electrical Engineering at
the University of New South Wales.

\newpage

\section*{FIGURE CAPTIONS}

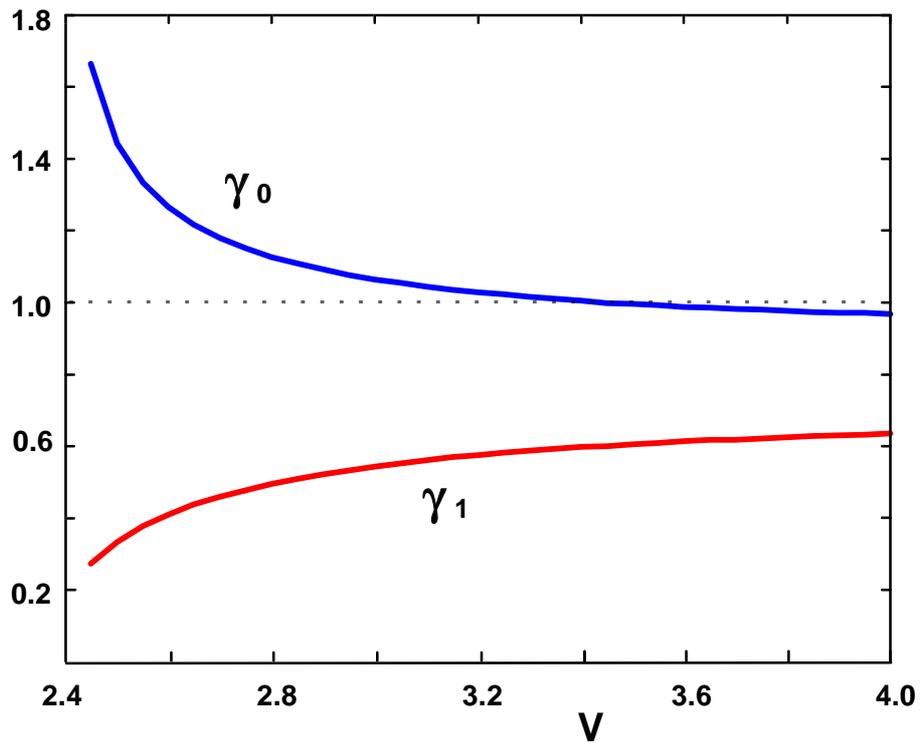
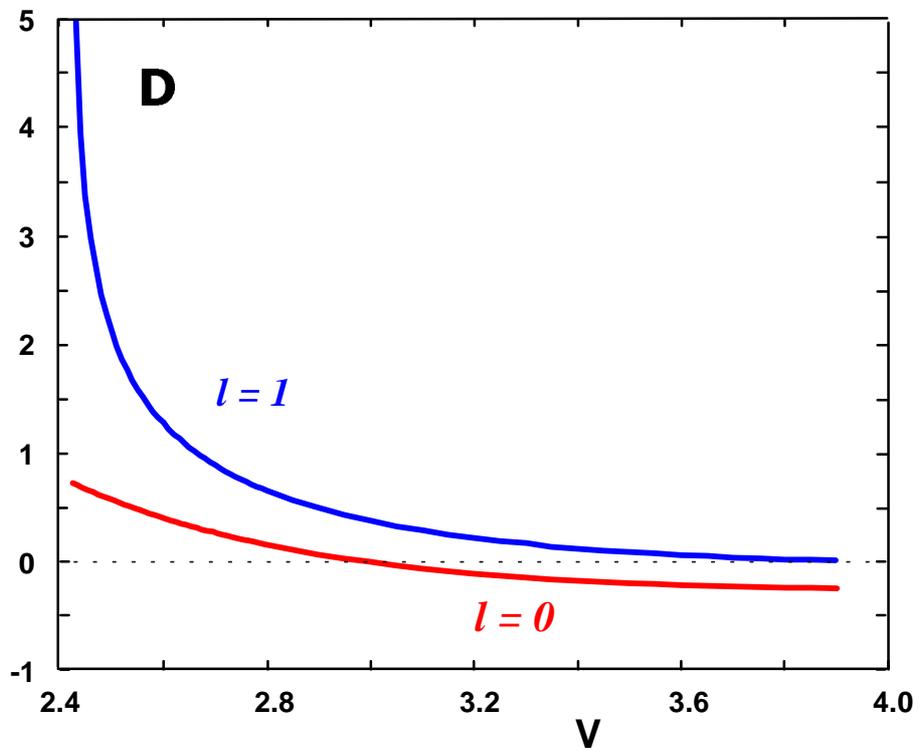
Fig. 1. The normalized XPM coefficients (a) and the normalized waveguide
dispersion coefficients (b), vs. the waveguide parameter $V$ for the
fundamental\ ($S=0$) and helical ($S=1$) modes in the standard step-index
fiber. The full waveguide dispersion is $\left( n_{{\rm cl}}V\Delta /\lambda
c\right) D$, where $c$ is the light velocity, and $\Delta \equiv (n_{{\rm co}%
}-n_{{\rm cl}})/n_{{\rm co}}$. In the plot (a), the dashed line shows the
usual value ($\equiv 1$) of the XPM coefficient.

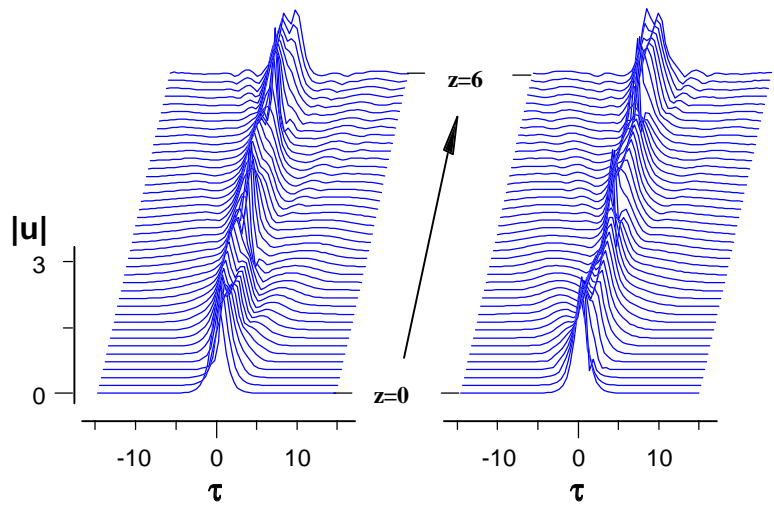
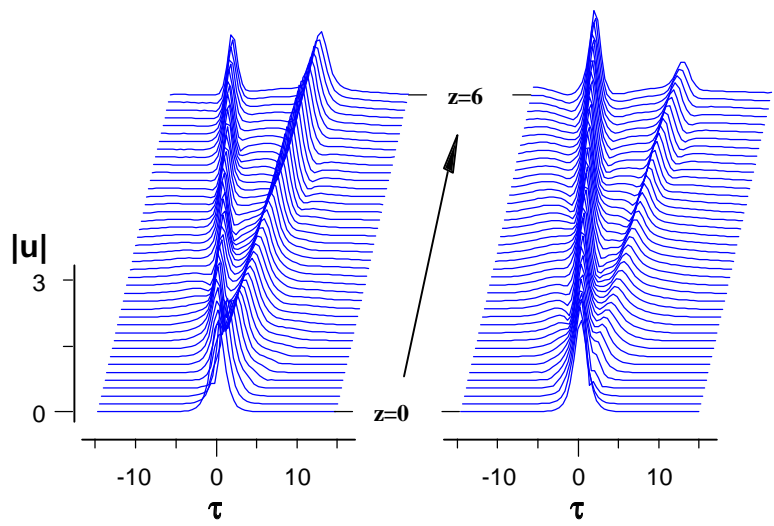
Fig. 2. Collisions of initially overlapped solitons with the amplitudes $%
|u|_{{\rm max}}=|v|_{{\rm max}}=2$ and the relative inverse group velocity $%
c_{1}-c_{0}=29$ (recall it must be a large parameter) simulated with Eqs.
(1) and (2), in which $\beta _{0}=\beta _{1}=-1$: (a) the usual WDM system,
with $\gamma _{0}=\gamma _{1}\equiv 1$; (b) the fundamental and helical
solitons, with $\gamma _{0}=0.98$ and $\gamma _{1}=0.62$. The simulations
were carried out from $z=0$ up to a distance equal to six soliton periods,
which is $z=6$.


\newpage

\begin{thebibliography}{99}
\bibitem{Agr}  G.P. Agrawal. {\it Nonlinear Fiber Optics} (Academic Press:
San Diego, 1995).

\bibitem{HK}  A. Hasegawa and Y. Kodama. {\it Solitons in Optical
Communications} (Oxford University Press: Oxford, 1995).

\bibitem{Love}  A.W. Snyder and J.D. Love. {\it Optical Waveguide Theory}
(Chapman and Hall: London, 1991)

\bibitem{Petrov}  D.V. Petrov, L. Torner, J. Martorell, R. Vilaseca, J.P.
Torres, and C. Cojocaru, Opt. Lett. {\bf 23}, 1444 (1998).

\bibitem{Elgin}  J.P. Silmon-Clyde and J.N. Elgin, Opt. Lett. {\bf 23}, 180
(1998).

\bibitem{coll}  S. Kumar, Y. Kodama, and A. Hasegawa, Electron. Lett. {\bf 33%
}, 459 (1997); A.N. Niculae, W. Forysiak, A.J. Gloag, J.H.B. Nijhof and N.J.
Doran, Opt. Lett. {\bf 23}, 1354 (1998); P.V. Mamyshev and L.F. Mollenauer,
Opt. Lett. {\bf 24}, 448 (1999).

\bibitem{Abl}  M.J. Ablowitz, G. Biondini, S. Chakravarty, and R.L. Horne,
Opt. Comm. {\bf 150}, 305 (1998).

\bibitem{KMJ}  D.J. Kaup, B.A. Malomed, and J. Yang, Opt. Lett. {\bf 23},
1600 (1998).

\bibitem{Silberberg}  Y. Barad and Y. Silberberg, Phys. Rev. Lett. {\bf 78},
3290 (1997).

\bibitem{spinning}  D. Mihalache, D. Mazilu, L.-C. Crasovan, B.A. Malomed,
and F. Lederer, Phys. Rev. E {\bf 61}, 7142 (2000).

\bibitem{hollow}  Y. Tamaki, K. Midorikawa, M. Obara, Appl. Phys. B, {\bf B67%
}, 59 (1998); Y. Tamaki, Y. Nagata, M. Obara, and K. Midorikawa, Phys. Rev.
A {\bf 59}, 4041 (1999); Z. Cheng, A. Furbach, S. Sartania, M. Lenzner, Ch.
Spielmann, and F. Krausz, Opt. Lett. {\bf 24}, 247 (1999).

\bibitem{Dan}  D. Anderson, Phys. Rev. A {\bf 27}, 3135 (1983).

\end{thebibliography}
\end{document}